\documentclass[letterpaper, 10 pt, conference]{ieeeconf}  % Comment this line out if you need a4paper

\usepackage[pdftex,dvipsnames]{xcolor}  % Coloured text etc.
\usepackage[colorinlistoftodos,prependcaption,textsize=tiny]{todonotes}

\IEEEoverridecommandlockouts                              % This command is only needed if you want to use the \thanks command
\usepackage{amsthm}
\usepackage{graphics} % for pdf, bitmapped graphics files
\usepackage{epsfig} % for postscript graphics files
\usepackage{times} % assumes new font selection scheme installed
\usepackage{amsmath} % assumes amsmath package installed
\usepackage{amssymb}  % assumes amsmath package installed

\usepackage[maxbibnames=99,giveninits=true]{biblatex}
\usepackage{bm}
\usepackage{xcolor}
\usepackage{graphicx}
\usepackage{algorithm}
\usepackage[noend]{algpseudocode}
\usepackage{booktabs}
\usepackage{siunitx}% An awesome package for typesetting and manipulation numbers and units.
\usepackage{hyperref}% Used for insert the link

\usepackage{caption}% Better control over caption
\usepackage{lipsum}% Example text
\usepackage{multirow}
\usepackage{diagbox} 
\usepackage{subcaption}

\newcommand{\R}{\mathbb{R}}%commands for easy math notations
\newcommand{\io}{\iota_{\varepsilon}}
\newcommand{\casr}{\mathcal{R}^{\mathcal{\bm{I}}_m,j}_{\varepsilon}}

\newcommand{\sigi}{\sigma_i}
\newcommand{\sigj}{\sigma_j}

\newtheorem{definition}{\bf Definition}
\newtheorem{assumption}{\bf Assumption}
\newtheorem{theorem}{\bf Theorem}
\newtheorem{lemma}{\bf Lemma}

\begin{document}
\title{\LARGE \bf
Cascading Waves of Fluctuation in Time-delay Multi-agent Rendezvous
}

\author{Guangyi Liu, Vivek Pandey, Christoforos Somarakis, and Nader Motee 
\thanks{
This work was supported by the ONR N00014-19-1-2478. \endgraf
G. Liu, V. Pandey, and N. Motee are with the Department of Mechanical Engineering and Mechanics, Lehigh University, Bethlehem, PA, 18015, USA. {\tt\small \{gliu,vkp219,motee\}@lehigh.edu}.\endgraf
 C. Somarakis is with the Intelligent Systems Lab, Palo Alto Research Center - a Xerox Company, Palo Alto, CA, 94304. {\tt\small somarakis@parc.com.}
}
}

\maketitle

% Replace this two for submission
\thispagestyle{plain}
\pagestyle{plain}

\begin{abstract} 
We develop a framework to assess the risk of cascading failures when a team of agents aims to rendezvous in time in the presence of exogenous noise and communication time-delay. The notion of value-at-risk (VaR) measure is used to evaluate the risk of cascading failures (i.e., waves of large fluctuations) when agents have failed to rendezvous. Furthermore, an efficient explicit formula is obtained to calculate the risk of higher-order cascading failures recursively. Finally, from a risk-aware design perspective, we report an evaluation of the most vulnerable sequence of agents in various communication graphs. 
\end{abstract}

%%%%%%%%%%%%%%%%%%%%%%%%%%%%%%%%%%%%%%%%%%%%%%%%%%%%%%%%%%%%%%%%%%%%%%%%%%%%%%%%%%%%%

\section{Introduction}
In the past decades, consensus networks have shown spectacular advantages in coordinating the networked control systems, e.g.,  \cite{krueger1998perception,fagiolini2008consensus}. However, like ``every coin has two sides", consensus networks or cyber-physical systems also show their vulnerability to large-scale failures \cite{bragagnolo2019attack,slay2007lessons,kushner2013real}. Given that most consensus networks suffer from communication or other physically induced limitations (e.g., time-delay), as well as external disturbances (e.g., biased opinion or noisy sensor measurements), individuals within the network will be inevitably driven away from the consensus and enter failure states. These failures are prone to cascade in the system and cause new failures to emerge \cite{liu2021risk,liu2022risk}; then, one needs to investigate how the existing failures are going to impact the rest of the system, i.e., the cascading (domino) effect, as well as developing systemic techniques to quantify such effects.

The domino effects in the systemic failure navigate our interests towards the cascading failures \cite{7438924, zhang2018cascading, zhang2019robustness}, and how their emergence will impact the system's safety. In other words, we aim to investigate how resilient the system is while {\it multiple} failures have already occurred in the system, as well as how likely the failure will cascade. Probing into the cascading failures and their emergence have many practical meanings from the system design perspective: It is essential to evaluate the performance of the networked control system under the impact of the existing failures, or, in other words, the ability to isolate the propagation of existing failures and keep the rest of the system safe.

In this paper, we will develop a theoretical framework based on systemic {\it risk analysis} \cite{rockafellar2000optimization,rockafellar2002conditional} to evaluate the possibility of cascading failures in the consensus network. Our objective is to utilize the risk quantification regime, e.g., the value-at-risk measure, to assess how likely the existing failures in the consensus network can trigger future failures throughout the system. Providing the quantification of such effects will give valuable insights into the networked control systems design and highlight the potential of a networked system to suffer from cascading systemic failures.

We consider the motivational scenario as a network of autonomous agents that aim to achieve rendezvous in time \cite{Somarakis2019g}. It is assumed that agents communicate over a time-invariant network with lagged transmission and processing of information from other agents. The network is also perturbed by statistical noise applied to each agent independently. The noise models the effect of external perturbations and turns the system into a stochastic dynamical network. We will focus on studying the event of a large fluctuation, i.e., failing to reach the rendezvous in time, given that some agents have already deviated from the consensus. 

{\it Our Contributions: }
As a natural outgrowth of our recent works on first-order linear consensus networks \cite{Somarakis2016g, Somarakis2017a, Somarakis2019g, liu2022risk}, this paper generalizes the concept of the cascading failures of two agents \cite{liu2021risk,liu2022risk} into the situation when there exists an arbitrary number of failures \cite{liu2022emergence}. First, we examine the value-at-risk measure of cascading failures in the steady-state distribution of observables for a time-delayed linear consensus network. In particular, explicit formulas are obtained for the risk of large fluctuation, given that other agents have failed to reach the consensus. Then, with extensive simulations, we evaluate the most vulnerable sequence of agents in a network, which is exclusively obtained with the generalized cascading risk for multiple existing failures. 

%%%%%%%%%%%%%%%%%%%%%%%%%%%%%%%%%%%%%%%%%%%%%%%%%%%%%%%%%%%%%%%%%%%%%%%%%%%%%%%%%%%%
\section{Mathematical Notation}
We denote the non-negative orthant of the Euclidean space $\R^n$ by $\mathbb{R}_{+}^n$, its set of standard Euclidean basis by $\{\bm{e}_1, \dots, \bm{e}_n\}$,  and  the vector of all ones by $\bm{1}_n = [1, \dots, 1]^T$. The absolute value operation for vectors is defined
element-wise, i.e., $|{\bm z}| = [|z_1|,..., |z_n|]^T$.

\vspace{0.1cm}
\noindent{\it Algebraic Graph Theory:} A weighted graph is defined by $\mathcal{G} = (\mathcal{V}, \mathcal{E}, \omega)$, where $\mathcal{V}$ is the set of nodes, $\mathcal{E}$ is the set of edges (feedback links), and $\omega: \mathcal{V} \times \mathcal{V} \rightarrow \mathbb{R}_{+}$ is the weight function that assigns a non-negative number (feedback gain) to every link. Two nodes are directly connected if and only if $(i,j) \in \mathcal{E}$.

\begin{assumption}  \label{asp:connected}
    Every graph in this paper is connected. In addition, for every $i,j \in \mathcal{V}$, the following properties hold:
    \begin{itemize}
        \item $\omega(i,j) > 0$ if and only if $(i,j) \in \mathcal{E}$.
        \item $\omega(i,j) = \omega(j,i)$, i.e., links are undirected.
        \item $\omega(i,i) = 0$, i.e., links are simple.
    \end{itemize}
    
\end{assumption}

The Laplacian matrix of $\mathcal{G}$ is a $n \times n$ matrix $L = [l_{ij}]$ with elements
\[
    l_{ij}:=\begin{cases}
        \; -k_{i,j}  &\text{if } \; i \neq j \\
        \; k_{i,1} + \ldots + k_{i,n}  &\text{if } \; i = j 
    \end{cases},
\]
where $k_{i,j} := \omega(i,j)$. The Laplacian matrix of a graph is symmetric and positive semi-definite \cite{van2010graph}. Assumption \ref{asp:connected} implies the smallest Laplacian eigenvalue is zero with algebraic multiplicity one. The spectrum of $L$ can be ordered as 
$
    0 = \lambda_1 < \lambda_2 \leq \dots \leq \lambda_n.
$
The eigenvector of $L$ corresponding to $\lambda_k$ is denoted by $\bm{q}_{k}$. By letting $Q = [\bm{q}_{1} | \dots | \bm{q}_{n}]$, it follows that $L = Q \Lambda Q^T$ with $\Lambda = \text{diag}[0, \lambda_2, \dots, \lambda_n]$. We normalize the Laplacian eigenvectors such that $Q$ becomes an orthogonal matrix, i.e., $Q^T Q = Q Q^T = I_{n}$ with $\bm{q}_1 = \frac{1}{\sqrt{n}} \bm{1}_n$. 

\vspace{0.1cm}

\noindent{\it Probability Theory:} Let $\mathcal{L}^{2}(\mathbb{R}^{q})$ be the set of all $\R^q-$valued random vectors $\bm{z} = [z^{(1)}, \dots ,z^{(q)}]^T$ of a probability space $(\Omega, \mathcal{F}, \mathbb{P})$ with finite second moments. A normal random variable $\bm{y} \in \mathbb{R}^{q}$ with mean $\bm{\mu} \in \mathbb{R}^{q}$ and $q \times q$ covariance matrix $\Sigma$ is represented by $\bm{y} \sim \mathcal{N}(\bm{\mu}, \Sigma)$. The error function $\text{erf}:\R \rightarrow (-1,1)$ is
$
\text{erf} (x) = \frac{2}{\sqrt{\pi}} \int_{0}^{x} e^{-t^2} \text{d} t,
$
which is invertible on its range as $\text{erf}^{-1} (x)$. We employ standard notation $\text{d} \bm{\xi}_t$ for the formulation of stochastic differential equations.

%%%%%%%%%%%%%%%%%%%%%%%%%%%%%%%%%%%%%%%%%%%%%%%%%%%%%%%%%%%%%%%%%%%%%%%%%%%%%%%%%%%%%
\section{Problem Statement}\label{problemstatement}
% \begin{figure}
%     \centering
% 	\includegraphics[width=\linewidth]{Figs/rendezvous_intro.jpg}
% 	\caption{ \guangyi{Change this figure and add caption}. }
% 	\label{fig:rendezvous}
% \end{figure}

We consider the class of time-delay linear consensus networks that have found wide applications in engineering, such as clock synchronization in sensor networks, rendezvous in space or time, and heading alignment in swarm robotics; we refer to \cite{ren2007information,olfati2007consensus} for more details. As a motivational application, we consider the rendezvous problem in time where the group objective is to meet {\it simultaneously} at a prespecified location known to all agents.\footnote{Rendezvous in space is very similar to rendezvous in time by switching the role of time and location.} Agents do not have a priori knowledge of the meeting time as it may have to be adjusted in response to unexpected emergencies or exogenous uncertainties \cite{ren2007information}. Thus, all agents should agree on a rendezvous time by achieving consensus. The consensus can be accomplished by each agent $i = 1, \dots, n$ creating a state variable, say $x_i \in \R$, representing its belief of the rendezvous time. Each agent's initial belief is set to its preferred time that it can rendezvous with others. Then, the rendezvous dynamics for each agent evolve in time according to the following stochastic differential equations:
\begin{equation}    \label{eq:dyn-one}
    \text{d}  x_t^{(i)} = u_t^{(i)} \, \text{d}  t + b \, \text{d}  \xi_t^{(i)},
\end{equation}
for all $i = 1, \dots, n$. Each agent's control input is $u_i(t) \in \R$. The source of uncertainty is diffused in the network as additive stochastic noise, and its magnitude is uniformly scaled by the diffusion coefficient $b \in \R$. The impact of uncertain environments on the dynamics of agents is modeled by independent Brownian motions $\xi_1, \dots, \xi_n$. In most real-world situations, agents experience a time-delay accessing, computing, or sharing their own and neighboring agents' state information \cite{ren2007information}. Hence, all agents are assumed to experience an identical time-delay  $\tau \in \R_{+}$. The control inputs are determined via a negotiation process by forming a linear consensus network over a communication graph using feedback law
\begin{equation}    \label{eq:feedback}
    u_t^{(i)} = \sum_{j = 1}^{n} k_{ij} \left(x_{t-\tau}^{(j)} - x_{t-\tau}^{(i)} \right),
\end{equation}
where $k_{ij} \in \R_{+}$ are non-negative feedback gains. Let us denote the state vector by $\bm{x}_t = [x_{t}^{(1)}, \dots, x_{t}^{(n)} ]^T$, and the vector of exogenous disturbance by $\bm{\xi}_t = [\xi_t^{(1)}, \dots, \xi_t^{(n)} ]^T$. The dynamics of the resulting closed-loop network can be cast as a linear consensus network that is governed by the time-delayed stochastic differential equation
\begin{equation}    \label{eq:dyn}
    \text{d}  \bm{x}_t = -L \, \bm{x}_{t-\tau}\, \text{d}  t + B \, \text{d}  \bm{\xi}_t
\end{equation}
for all $t \geq 0$, where the initial function $\bm{x}_t=\phi(t)$ is deterministically given for $t \in [-\tau, 0]$ and $B = b I_n$. The underlying coupling structure of the consensus network \eqref{eq:dyn} is a graph $\mathcal{G}$ that satisfies Assumption \ref{asp:connected} with Laplacian matrix $L$. It is considered that the communication graph $\mathcal{G}$ is time-invariant such that the network of agents aim to reach a consensus on a rendezvous time before they perform motion planning to get to the meeting location. Upon reaching a consensus, a properly designed internal feedback control mechanism steers each agent toward the rendezvous time.

\begin{assumption}  \label{asp:stable}
    The time-delay satisfies $\tau < \frac{\pi}{2 \lambda_n}$.
\end{assumption}

When there is no noise, i.e., $b = 0$, it is known \cite{olfati2004consensus} that under Assumption \ref{asp:connected} and \ref{asp:stable}, states of all agents converge to the average of all initial states $\frac{1}{n} \bm{1}_n^T \bm{x}_0$; whereas in the presence of input noise, state variables fluctuate around the average value $\frac{1}{n} \bm{1}_n^T \bm{x}_t$. To quantify the quality of rendezvous and its fragility features, we consider the vector of observables 
\begin{equation} \label{eq:observables}
     \bm{y}_t = M_n \, \bm{x}_t,
\end{equation}
in which $M_n = I_n - \frac{1}{n} \bm{1}_n \bm{1}_n^T $ is the centering matrix and the observable $\bm{y}_t = (y_t^{(1)}, ..., y_t^{(n)})^T$ measures the agents' deviations from consensus. Assumption \ref{asp:connected} implies that one of the modes of network \eqref{eq:dyn} is marginally stable. The marginally stable mode, which corresponds to the zero eigenvalue of $L$, is unobservable from the output \eqref{eq:observables}, which keeps $\bm{y}_t$ bounded in the steady-state. When the noise is absent, we have $\bm{y}_t \rightarrow 0$ as $t \rightarrow \infty$. Consequently, the exogenous noise excites the observable modes of the network, and the output fluctuates around zero. This implies that agents will not agree upon an exact rendezvous time. A practical resolution is to allow a tolerance interval for agents to concur.

\begin{definition}
    For a given tolerance level $c \in \R_+$, the consensus network \eqref{eq:dyn} reaches $c$-consensus if it can tolerate some degrees of disagreement such that
    \begin{equation} \label{eq:event}
        \lim_{t \rightarrow \infty} |\bm{y}_t| \leq c \, \bm{1}_n
    \end{equation}
    holds with high probability\footnote{The high probability means a probability larger than a predefined cut-off number close to one.}.
\end{definition}

The notion of $c$-consensus means that all agents have agreement on all points in $\{\bm{x} \in \R^n \,\big|\,  |M_n \bm{x}| \leq c \bm{1}_n\}$. Suppose that event \eqref{eq:event} holds, then the network of agents will achieve a $c$-consensus over the rendezvous time in the following sense. In steady-state, the $i$'th agent is assured that by $x_t^{(i)} \pm \,c$ units of time, all other agents will arrive and meet each other in that time interval with high probability. At the same time, some undesirable situations that we refer to as systemic failures may also happen.

\begin{definition}
    For a given tolerance level $c \in \R_+$, an agent with observable $y_t^{(i)}$ is prone to systemic failure, i.e., the large fluctuation, if the probability of the  stochastic event
    \begin{equation} \label{eq:failure}
        |y_t^{(i)}| > c
    \end{equation}
    is nonzero as $t \rightarrow \infty$.
\end{definition}

The {\it problem} is to quantify the risk of large fluctuations conditioned on the knowledge that a subset of agents has already deviated from the consensus. We refer to this quantity as the cascading risk of large fluctuations, and our goal is to characterize its value as a function of the underlying communication graph topology, time-delay, and statistics of noise. To this end, we will develop a general framework to assess the cascading risk of large fluctuations using the steady-state statistics of the closed-loop system \eqref{eq:dyn}.

The rest of the paper is organized as follows. First, in \S \ref{prelims}, we present a few key preliminary results on the steady-state behavior and statistics of the noisy consensus network. These results help us shape the closed-form risk formula of cascading failures in \S \ref{sec:risk}, which constitutes the main contribution of this work. Finally, in \S \ref{sec:update}, we also present an update law that reveals the evolution of the cascading risk when new failures are discovered in the system. All proofs of theoretical results are shown in the appendix of this paper.

% Next, the special graph topology is examined in \S \ref{specialgraph}, where we present the complexity of cascading failure phenomena and the time-delay induced fundamental limits for the network design. 

%%%%%%%%%%%%%%%%%%%%%%%%%%%%%%%%%%%%%%%%%%%%%%%%%%%%%%%%%%%%%%%%%%%%%%%%%%%%%%%%%%%%%
\section{Preliminary Results}\label{prelims}

In this section, we present the steady-state statistics of network observables under external disturbances and time-delay. Such statistics form the foundation for the latter results on the value-at-risk measure.

\subsection{Steady-State Statistics of Observables}
Let us denote the steady-state value of the observable $y_i(t)$ by $\bar{y}_i := \lim_{t\rightarrow +\infty} y_t^{(i)}$ whenever it exists. It is known from \cite{Somarakis2019g} that if Assumption \ref{asp:connected} and \ref{asp:stable} are satisfied, as $t \rightarrow \infty$, the statistics of steady-state observables \eqref{eq:observables} follows a normal distribution  $\bm{\bar{y}} \sim \mathcal{N}(0, \Sigma)$, where $\Sigma$ is a $n \times n$ covariance matrix .
\begin{lemma}     \label{lem:y_steady}
    The elements of $\Sigma = [\sigma_{ij}]$ are given by
    \begin{equation} \label{eq:sigma_y}
    \begin{aligned}
        \sigma_{ij} = 
        \frac{1}{2} b^2 \sum_{k=2}^{n} \frac{\cos (\lambda_k \tau)}{\lambda_k (1 - \sin (\lambda_k \tau))} (\bm{m}_i^T \bm{q}_k)(\bm{m}_j^T \bm{q}_k),
    \end{aligned}
    \end{equation}
    where $\bm{m}_i$ denotes the $i$'th column of $M_n$ for all $i,j = 1,\dots, n$.  
\end{lemma}
For simplicity, we write the diagonal elements $\sigma_{ii}$ as $\sigi^2$. One can also find out that the cross-correlation $\rho_{ij} = \frac{\sigma_{ij}}{\sigi \sigj}$ is independent of $b$, the magnitude of external noises. This observation implies that the probabilistic interrelation between two agents originates from the time-delay and the communication graph but not the magnitude of the stochastic perturbation.

\subsection{Value-at-Risk Measure}
To quantify the uncertainty level encapsulated in agents' observables, we employ the notion of Value-at-Risk (VaR) measure \cite{Follmer2016,rockafellar2000optimization,Somarakis2016g,Somarakis2020b}. The VaR indicates the chance of a random variable landing inside an undesirable set of values that characterizes a systemic failure. The set of such undesirable values is referred to as a systemic set and denoted by $U \subset \R$. 
In probability space $(\Omega, \mathcal{F}, \mathbb{P})$, the set of systemic events of random variable $y: \Omega \rightarrow \R$ is defined as $\{ \omega \in \Omega ~|~y(\omega) \in U\}$. We define a collections of supersets $\{U_{\delta}~|~\delta \in [0,\infty]\}$ of $U$ that satisfy the following conditions for any sequence $\{\delta_n\}^{\infty}_{n=1}$ with property $\lim_{n \rightarrow \infty} \delta_n = \infty$ 
% \vivek{(We define a collections of super-sets $\{U_{\delta}~|~\delta \in [0,\infty]\}$, for any sequence $\{\delta_n\}^{\infty}_{n=1}$ with property $\lim_{n \rightarrow \infty} \delta_n = \infty$, that satisfy the following conditions  )}
\begin{itemize}
    \item $U_{\delta_2} \subset U_{\delta_1}$ when $\delta_1 < \delta_2$
    \item $\lim_{n \rightarrow \infty} U_{\delta_n} = \bigcap_{n=1}^{\infty} U_{\delta_n} = U$.
\end{itemize}

\begin{figure}[t]
    \centering
	\includegraphics[width=\linewidth]{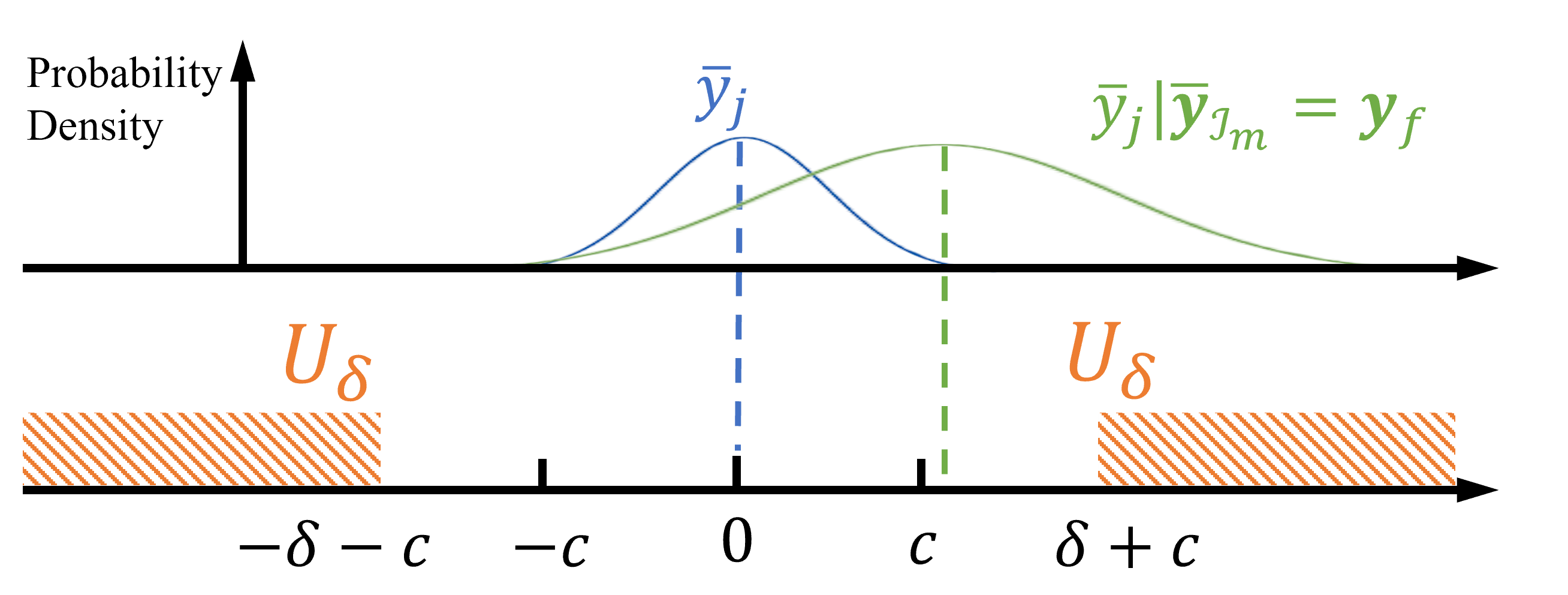}
	\caption{This figure depicts the concept of the risk set $U_{\delta}$ and how the distribution of $\bar{y}_j$ changes when $\bar{\bm{y}}_{\mathcal{I}_m} = \bm{y}_{f}$ is introduced.}
	\label{fig:risk-set}
\end{figure}

In practice, one can tailor the super-sets to cover a suitable neighborhood of $U$ to characterize alarm zones as a random variable approaches $U$. For a given $\delta >0$, the chance of $\{ y \in U_{\delta}\}$ indicates how close $y$ can get to $U$ in probability. For a given design parameter $\varepsilon \in (0,1)$, the VaR measure $\mathcal{R}_{\varepsilon} : \mathcal{F} \rightarrow \R_{+}$ is defined by
\begin{equation*}
    \mathcal{R}_{\varepsilon} := \inf \left\{ \delta>0 ~\big|~ \mathbb{P} \left\{y \in U_{\delta} \right\} < \varepsilon \right\},
\end{equation*}
where a smaller $\varepsilon$ indicates a  higher level of confidence on random variable $y$ to stay away from $U_\delta$. Let us elaborate and interpret what typical values of $\mathcal{R}_{\varepsilon}$ imply. The case $\mathcal{R}_{\varepsilon}=0$ signifies that the probability of observing $y$ dangerously close to $U$ is less than $\varepsilon$. We have $\mathcal{R}_{\varepsilon} > 0$ if and only if $y \in U_{\delta}$ for  $\delta >  \mathcal{R}_{\varepsilon}$ with probability less  than $\varepsilon$. 
In addition to several interesting properties (see for instance \cite{artzner1997thinking,artzner1999coherent,Somarakis2020b}), the risk measure is non-increasing with $\varepsilon$. We refer to Fig. \ref{fig:risk-set} for an example.

%%%%%%%%%%%%%%%%%%%%%%%%%%%%%%%%%%%%%%%%%%%%%%%%%%%%%%%%%%%%%%%%%%%%%%%%%%%%%%%%%%%%%
\section{Risk of Cascading Failures}   \label{sec:risk}

As a natural outgrowth of our previous work \cite{liu2022risk}, we generalize the definition of the ``cascading failures" from assuming only one existing failure into an arbitrary number of existing failures \cite{liu2022emergence}. To this end, let us consider the risk of cascading large fluctuation for the $j$'th agent, assuming that agents with ordered indices $\mathcal{I}_m =\{i_1, \cdots, i_m\}$ for some $m < n$ have failed to reach the $c-$consensus, where $j \notin \mathcal{I}_m$ and $i_1<i_2<...<i_m$. To evaluate the effect of the prior large fluctuations on the $j$'th agent, let us form a $(m+1) \times (m+1)$ block covariance matrix 
\begin{equation}    \label{eq:block_cov}
    \tilde{\Sigma} = \begin{bmatrix}\,
       \tilde{\Sigma}_{11} &\tilde{\Sigma}_{12}\\
       \tilde{\Sigma}_{21} &\tilde{\Sigma}_{22}\,
    \end{bmatrix},
\end{equation}
 where
$
\tilde{\Sigma}_{11} = \sigma_{j}^2, \, \tilde{\Sigma}_{12} = \tilde{\Sigma}_{21}^T = [\sigma_{j i_1},...,\sigma_{j i_m}]$, and  $\tilde{\Sigma}_{22} = [\sigma_{k_1 k_2}]_{k_1, k_2 \in \mathcal{I}_m} \in \R^{m \times m}$. The values of $\sigma_{ij}$s are computed using the result from \eqref{eq:sigma_y}. 

We also denote that observables of the failed agent are represented by vector $\bm{y}_{f} = [y_{f_1}, ... , y_{f_m}]^T$. For instance, when the $i$'th agent failed to each the $c-$consensus, one may set the value of $y_{f_i}$ to satisfy $|y_{f_i}| > c$. To obtain the statistics of the observable when multiple failures occur, one should calculate the steady-state conditional probability distribution of $\bar{y}_j$ given 
$\bm{\bar{y}}_{\mathcal{I}_m}  = \bm{y}_{f}$ 
within a multivariate normal distribution, where $\bm{\bar{y}}_{\mathcal{I}_m}=[\bar{y}_{i_1}, ... , \bar{y}_{i_m}]^T$.

\begin{lemma}   \label{lem:multi-condition}
    Suppose that the system \eqref{eq:dyn} reaches the steady-state, the conditional distribution of $\bar{y}_j$ given $\bm{\bar{y}}_{\mathcal{I}_m} = \bm{y}_{f}$ follows a normal distribution $\mathcal{N}(\tilde{\mu},\tilde{\sigma}^2)$ with 
   \begin{equation} \label{eq:lem2}
       \tilde{\mu} = \tilde{\Sigma}_{12}\tilde{\Sigma}_{22}^{-1}\bm{y}_f, 
    \text{ and }
    ~ \tilde{\sigma}^2 = \tilde{\Sigma}_{11} -  \tilde{\Sigma}_{12}\tilde{\Sigma}_{22}^{-1}\tilde{\Sigma}_{21},
   \end{equation}
    where the sub-blocks $\tilde{\Sigma}_{11}, ..., \tilde{\Sigma}_{22}$ are defined in \eqref{eq:block_cov}.
\end{lemma}

The above lemma reveals the statistics of the observables when there exist multiple large fluctuations (failures) in the consensus network. In this scenario, we aim to evaluate the risk of cascading failure on the $j$'th agent. The corresponding event is shown by 
\begin{equation*}
    \left\{ \bar{y}_{j} \in U_{\delta} ~ | ~ \bar{\bm{y}}_{\mathcal{I}
    _m}= \bm{y}_f \right\}, 
\end{equation*}
where the family of risk set\footnote{It should be noted that in our definition of $U_\delta$ for $c$-consensus, the family of supersets obviously depends on $c$ but its dependency to this parameter is omitted from $U_\delta$ for notational simplicity.} of cascading failure for the $j$'th agent is given by
\begin{equation*}
    U_{\delta}=\left( -\infty, -\delta - c\right) \bigcup \left(  \delta+c  ,\infty \right),
\end{equation*}
with $c \in \R_{+}$ and $\delta \in [0,\infty)$. Then, the value-at-risk measure for the cascading failure on the $j$'th agent is defined as
\begin{align}\label{eq:var}
    \casr = \inf \big\{ \delta > 0 \;\big|\; \mathbb{P} \{ \bar{y}_{j} \in U_{\delta} \;\big|\; \bar{\bm{y}}_{\mathcal{I}
    _m}= \bm{y}_f\} < \varepsilon \big\},
\end{align}
where $\varepsilon$ takes value in $(0,1)$. The cascading risk $\casr$ takes value in $[0,\infty)$, and it measures the chance of an agent failing to reach the $c-$consensus given that $m$ other agents have failed to rendezvous.

For the exposition of the next result, we introduce the following notation:
\begin{align}   \label{eq:kappa}
\kappa_{\delta,\pm}^{\mathcal{I}_m,j} = \frac{(\delta+c) \pm \tilde{\mu} }{\sqrt{2}\tilde{\sigma}},
\end{align}
with $ \tilde{\mu}$ and $\tilde{\sigma}$ defined in Lemma \ref{lem:multi-condition}. The notation below is also introduced for simplicity,
\begin{align} \label{eq:S}
    S(\delta) = \inf \left\{ \delta > 0 \, \bigg|\, \text{erf}(\kappa_{\delta,+}^{\mathcal{I}_m,j}) + \text{erf}(\kappa_{\delta,-}^{\mathcal{I}_m,j}) > 2 (1- \varepsilon) \right\}.
\end{align}

\begin{theorem} \label{thm:cas_ren_risk}
    Suppose that the consensus network \eqref{eq:dyn} reaches the steady-state and $m$ agents with labels $\mathcal{I}_m$ have failed to reach the $c$-consensus with the vector of observable $\bar{\bm{y}}_{\mathcal{I}_m} = \bm{y}_f$. The risk of cascading large fluctuation of agent $j$ is 
    \[
        \casr:=\begin{cases}
            0, &\text{if} ~ 1 - \frac{1}{2}\left(\text{erf}(\kappa_{0,+}^{\mathcal{I}_m,j}) + \text{erf}(\kappa_{0,-}^{\mathcal{I}_m,j}) \right) \leq \varepsilon \\
            S(\delta), &\text{otherwise} 
            \end{cases}
    \]
    where $j = 1, \dots, n$, $j \notin \mathcal{I}_m$, $|\bm{y}_{f}| > c\, \bm{1}_m$, and $S(\delta)$ is defined by \eqref{eq:S}.
\end{theorem}

The two cases of the cascading risk value are self-explanatory. The case $\casr = 0$ indicates the scenario when the probability of the $j$'th agent failing to reach $c-$consensus is always less than $\varepsilon$, which commonly corresponds to a low confidence level or the conditional distribution that lands away from the undesired set of the observable values. In the other case, the risk obtains a positive value based on the solution of \eqref{eq:S}. A higher value of $\casr$ usually indicates a higher chance that the cascading failure will occur in the system. The value of $c$ will also play the role as a design parameter when $\casr = 0$ since the system will reach the $c-$consensus with the confidence level $1-\varepsilon$. 

When the existing failures are not correlated to the $j$'th agent, i.e., $\tilde{\Sigma}_{12} = \tilde{\Sigma}_{21}^T = \bm{0}$, the cascading risk expression $\casr$ boils down to the risk of large fluctuation for a single agent to rendezvous, 
\begin{equation*} 
    \mathcal{R}^{j}_{\varepsilon} = \sqrt{2} \, \sigj \io  -c, \text{ if } \sigma_{j} > \dfrac{c} {\sqrt{2}\io},
\end{equation*}
where $\io = \text{erf}^{-1}(1- \varepsilon)$. This formula expands the risk of large fluctuations expression derived in  \cite{Somarakis2019g}, which considers deviations on $c$-consensus state for $c=0$.

Given that $m$ agents have failed to reach the $c$-consensus, one can formulate the risk profile of the network by stacking up all the cascading risk $\casr$ among remaining agents with a systemic risk profile vector $\bm{\mathcal{R}}^{\mathcal{I}_m}_{\varepsilon} \in \R^{n}$  as follows
\begin{equation*}
    \bm{\mathcal{R}}^{\mathcal{I}_m}_{\varepsilon} = \big[\mathcal{R}^{\mathcal{I}_m,1}_{\varepsilon},\dots , \mathcal{R}^{\mathcal{I}_m,n}_{\varepsilon} \big]^T,
\end{equation*}
where $\casr = 0$ if $j \in \mathcal{I}_m$.

\section{The Update Law for the Cascading Risk Computation}   \label{sec:update}
\begin{figure*}[t]
    \begin{subfigure}[t]{.19\linewidth}
        \centering
    	\includegraphics[width=\linewidth]{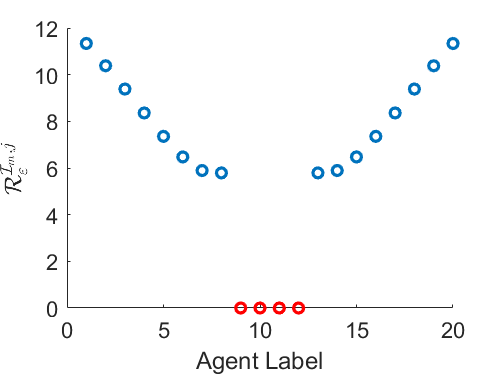}
    	\caption{The path graph.}
    \end{subfigure}
    \hfill
    \begin{subfigure}[t]{.19\linewidth}
        \centering
    	\includegraphics[width=\linewidth]{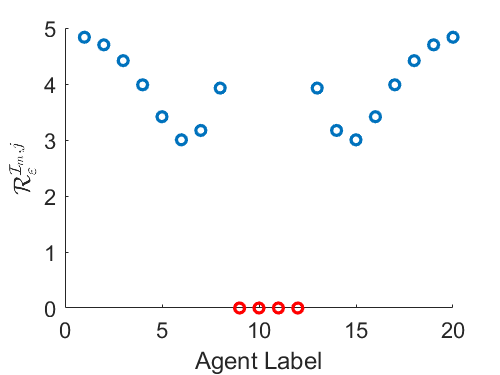}
    	\caption{The $2-$cycle graph.}
    \end{subfigure}
    \hfill
    \begin{subfigure}[t]{.19\linewidth}
        \centering
    	\includegraphics[width=\linewidth]{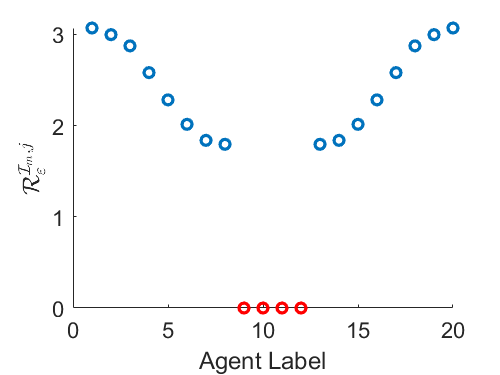}
    	\caption{The $5-$cycle graph.}
    \end{subfigure}
    \hfill
    \begin{subfigure}[t]{.19\linewidth}
        \centering
    	\includegraphics[width=\linewidth]{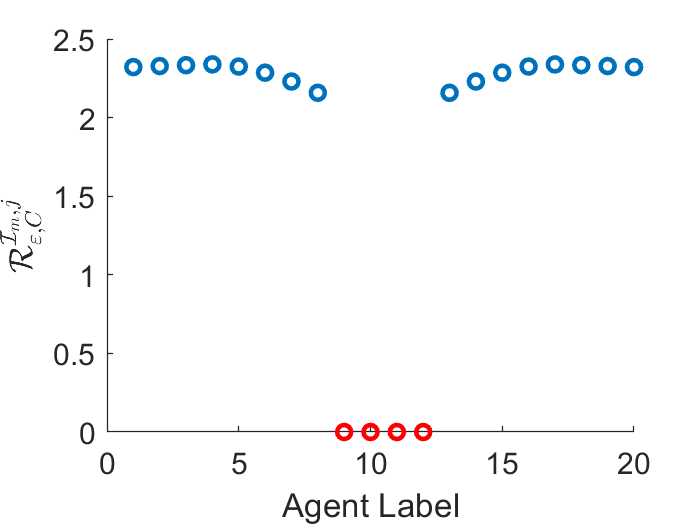}
    	\caption{The $7-$cycle graph.}
    \end{subfigure}
    \hfill
    \begin{subfigure}[t]{.19\linewidth}
        \centering
    	\includegraphics[width=\linewidth]{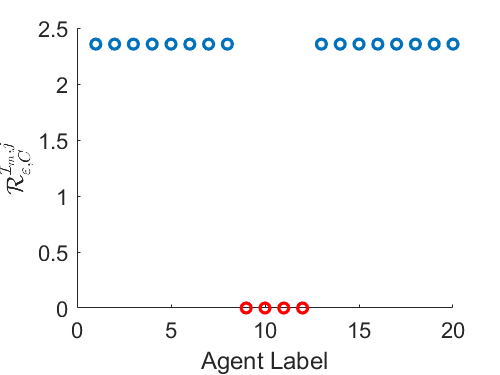}
    	\caption{The complete graph.}
    \end{subfigure}
    \caption{The risk profile of cascading failure  $\mathcal{R}^{\mathcal{I}_{m}}_{\varepsilon}$ when the existing large fluctuations have occurred at agents $\{9,10,11,12\}$.}
    \label{fig:cas_risk}
\end{figure*}

% \guangyi{Reveal how the cascading will evolve w.r.t the current risk value.}

% \guangyi{Vivek: Please help to edit this section with reasons about how the update law will reveal the evolution of risks.}

% Evaluating the cascading risk in a network requires the computation of the conditional distribution of $\bar{y}_j$. Based on the result of Lemma \ref{lem:multi-condition}, the observables of the existing failures, i.e., $\bm{y}_f$, is necessary for the computation of $\bar{y}_j$. 
% Sometimes one does not enjoy the luxury of possessing the full information of the value of $\bm{y}_f$ but still wants to explore how the risk of cascading failures will react when new failures are observed in the system.
% \guangyi{Saying why this situation is common.} Thus, 

% In this section,
% we consider the situation when $m$ failures have occurred in the system, and one only has the knowledge of the system parameters and conditional distribution of $\bar{y}_j$, but not the value of existing failures $\bar{\bm{y}}_{\mathcal{I}_m} = \bm{y}_f$. In this situation, we still aim to quantify the cascading risk on the $j$'th agent when a new failure is introduced into the system.

% To resolve this issue,
In this section, let us consider the scenario where agents with labels $\mathcal{I}_m$ are found in failure states, and we aim to update the statistics of the agent of interest, i.e., $\bar{y}_{j}|\bar{\bm{y}}_{\mathcal{I}_m}= \bm{y}_f$, when a new failure at agent $k|_{k \notin \mathcal{I}_m}$ is discovered. For the exposition of the next result, let us introduce the following notations 
\begin{equation}    \label{eq:j_and_k}
    \begin{aligned}
        &\tilde{\mu}_j = \tilde{\Sigma}_{12}(j)\tilde{\Sigma}_{22}^{-1}\bm{y}_f,
        ~~ \tilde{\sigma}^2_j = \sigj^2 -  \tilde{\Sigma}_{12}(j)\tilde{\Sigma}_{22}^{-1}\tilde{\Sigma}_{21}(j),\\
        &\tilde{\mu}_k = \tilde{\Sigma}_{12}(k)\tilde{\Sigma}_{22}^{-1}\bm{y}_f,
        ~~ \tilde{\sigma}^2_k = \sigma_k^2 -  \tilde{\Sigma}_{12}(k)\tilde{\Sigma}_{22}^{-1}\tilde{\Sigma}_{21}(k),\\
    \end{aligned}
\end{equation}
where the value of $\sigj, \sigma_k$ are computed using \eqref{eq:sigma_y}, the values of $\tilde{\Sigma}_{12}(k) = \tilde{\Sigma}_{21}^T(k)$ are given by \eqref{eq:block_cov} in the view of the agent $k$ \footnote{The values of $\tilde{\Sigma}_{12}(j) = \tilde{\Sigma}_{21}^T(j)$ are obtained with agent $j$, respectively.}, and the matrix $\tilde{\Sigma}_{22}^{-1}$ is computed using \eqref{eq:block_cov}. 

\begin{theorem}   \label{thm:conditional_prob_update}
    Suppose that $\bar{y}_j$ follows  $\mathcal{N}(\tilde{\mu}_j,\tilde{\sigma}_j^2)$ when $m$ agents have already failed with labels $\mathcal{I}_m$. The updated  
    conditional distribution of $\bar{y}_{j}$ when a new agent fails, i.e., agent $k \notin \mathcal{I}_m$ with observable $|y_{f_k}| > c$, is given by $\mathcal{N}(\tilde{\mu}',\tilde{\sigma}'^{2})$, such that
    \begin{align}   \label{eq:update}
        \tilde{\mu}' = \tilde{\mu}_j - \frac{\tilde{\sigma}_{jk}}{\tilde{\sigma}_{k}^2} (\tilde{\mu}_k - y_{f_k}),
        \text{ and }
        ~ \tilde{\sigma}'^{2} = \tilde{\sigma}_{j}^2 -  \frac{\tilde{\sigma}_{jk}}{\tilde{\sigma}_k^{2}},
    \end{align}
    where 
    $$
    \tilde{\sigma}_{jk} = \sigma_{jk} - \tilde{\Sigma}_{12}(j) \tilde{\Sigma}_{22}^{-1} \tilde{\Sigma}_{21}(k),
    $$
    and the value of $\tilde{\mu}_j, \tilde{\mu}_k,  \tilde{\sigma}_j, \tilde{\sigma_{k}}$ are computed using \eqref{eq:j_and_k}.
\end{theorem}

The above result provides a good insight into the change in the conditional mean and covariance of an agent when the number of failed agents increase from $m$ to $m + 1$. Interestingly, the conditional covariance of the agent $j$ is independent of the value of the deviation of the failed agent $k$. However, the conditional mean of the agent $j$ depends on the difference between the conditional mean of the failed agent $k$, when $m$ failures have occurred, and its current value of the failed observable, $y_{f_k}$. In the unlikely case of $y_{f_k} = \tilde{\mu}_k$, the conditional mean of the agent $j$ remains unchanged.

In addition, the above result can be applied to find the sequence of agents with maximum risk after each additional failure. This information can be very helpful in designing an optimal communication network to protect a sequence of agents that is most vulnerable to cascading systemic failures.

% The above result provides an update rule for the conditional distribution when any additional failure occurs \vivek{This statement can be omitted}. \vivek{This part needs to be edited or completely removed if we cannot prove the computational efficiency}The update law \eqref{eq:update} only requires the computation of the inversion $\tilde{\Sigma}_{22}^{-1}$, which is already known from the previous step, thus helping to avoid the matrix inverse calculations at each step. One can use this update law to compute the risk for all $m = 1,..., \bar{m}$ in a computationally efficient manner to obtain the conditional distributions and their corresponding risks.

%%%%%%%%%%%%%%%%%%%%%%%%%%%%%%%%%%%%%%%%%%%%%%%%%%%%%%%%%%%%%%%%%%%%%%%%%%%%%%%%%%%%%
\section{Case Studies}
We discuss the case studies for the rendezvous problem with the consensus dynamics governed by \eqref{eq:dyn} among the complete, path, and the p-cycle communication graphs \cite{van2010graph}. In all case studies, we consider agents with labels $\mathcal{I}_m$ cannot achieve the $c$-consensus and have been found with large fluctuations $\bm{y}_f = y_f\bm{1}_m$. Let us consider $n = 20, c = 0.1, y_f = 2, b = 4, \tau = 0.05 \text{ and } \varepsilon = 0.1$ for all simulations.

\subsection{Risk of Cascading Large Fluctuation}

The risk profile of the cascading failure to reach $c$-consensus $\bm{\mathcal{R}}^{\mathcal{I}_m}_{\varepsilon}$ is computed using the closed-form expression derived in Theorem \ref{thm:cas_ren_risk} for various unweighted communication graph topologies, which is shown in Fig. \ref{fig:cas_risk}. There are a few interesting results worth reporting: In path graph \cite{van2010graph}, it is shown that the riskiest agent is found at both ends of the graph when there exist multiple failures in the system; In $p-$cycle graphs \cite{van2010graph}, one can observe that the risk profile pattern converges to the complete graph as $p$ increases. Finally, the complete graph shows that the risk value remains unchanged regardless of the agent of the interest.

\subsection{Characteristics of Existing Failures}
\begin{figure*}[t]
    \begin{subfigure}[t]{.24\linewidth}
        \centering
    	\includegraphics[width=\linewidth]{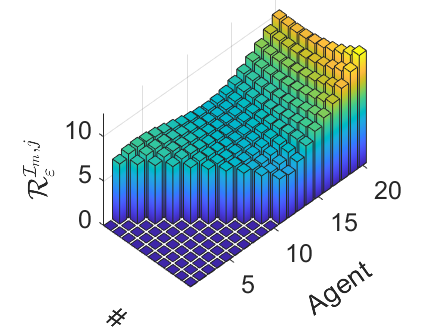}
    	\caption{The path graph.}
    \end{subfigure}
    \hfill
    \begin{subfigure}[t]{.24\linewidth}
        \centering
    	\includegraphics[width=\linewidth]{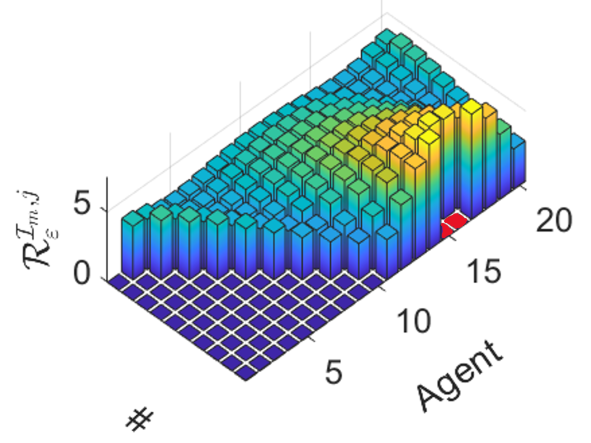}
    	\caption{The $2-$cycle graph.}
    \end{subfigure}
    \hfill
    \begin{subfigure}[t]{.24\linewidth}
        \centering
    	\includegraphics[width=\linewidth]{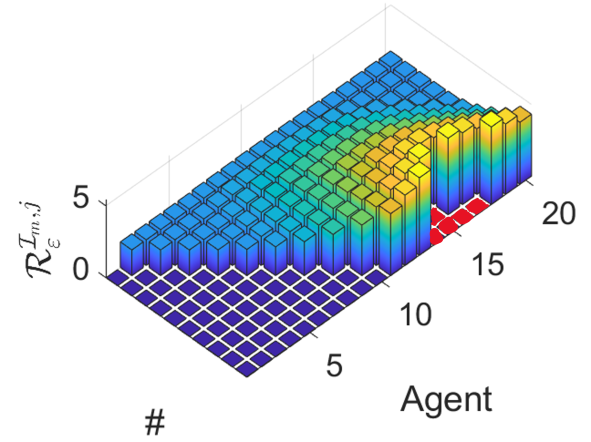}
    	\caption{The $5-$cycle graph.}
    \end{subfigure}
    \hfill
    \begin{subfigure}[t]{.24\linewidth}
        \centering
    	\includegraphics[width=\linewidth]{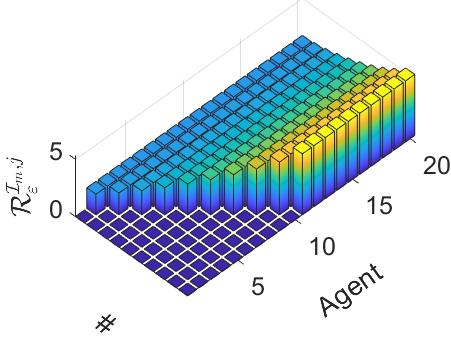}
    	\caption{The complete graph.}
    \end{subfigure}
    \hfill
    \begin{subfigure}[t]{.24\linewidth}
        \centering
    	\includegraphics[width=\linewidth]{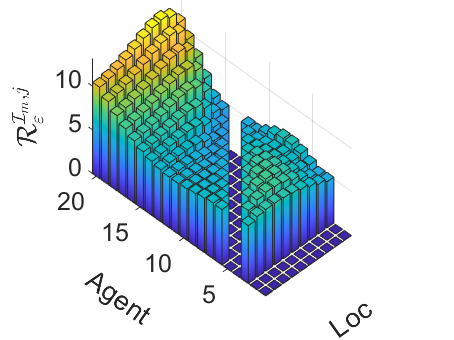}
    	\caption{The path graph.}
    \end{subfigure}
    \hfill
    \begin{subfigure}[t]{.24\linewidth}
        \centering
    	\includegraphics[width=\linewidth]{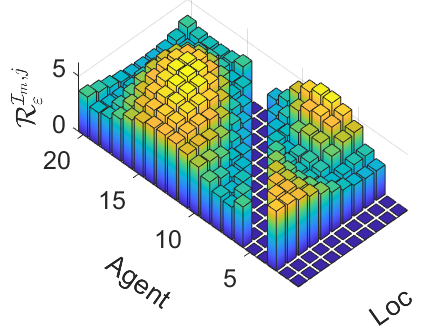}
    	\caption{The $2-$cycle graph.}
    \end{subfigure}
    \begin{subfigure}[t]{.24\linewidth}
        \centering
    	\includegraphics[width=\linewidth]{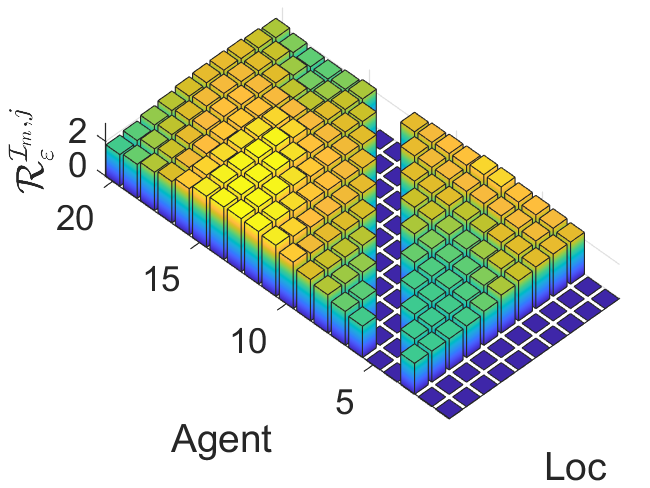}
    	\caption{The $5-$cycle graph.}
    \end{subfigure}
    \hfill
    \begin{subfigure}[t]{.24\linewidth}
        \centering
    	\includegraphics[width=\linewidth]{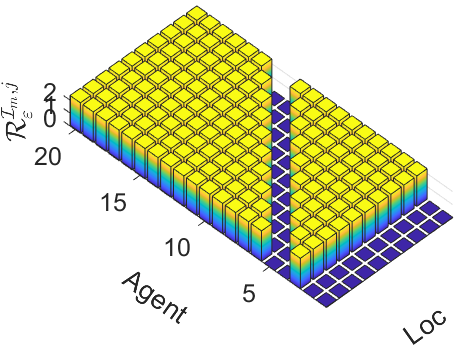}
    	\caption{The complete graph.}
    \end{subfigure}
    
    \caption{The cascading risk profile with (a)-(d) a different number of failures; (e) - (h) various spatial distributions. The agents that obtain the risk value as $0$ represent existing failures; the red color denotes when the risk value reaches $+\infty$.}
    \label{fig:risk_cha}
\end{figure*}

In the cascading risk, the characteristics of the existing failures will profoundly affect the cascading risk profile. For instance, the number of existing failures and their spatial distribution in the network reshapes the cascading risk profile in various networks. This phenomenon is examined in Fig. \ref{fig:risk_cha}, in which a different number of failures and various failures distribution are considered. Some of these results can be validated with our theoretical findings, e.g., in a complete graph, the risk profile obtains a constant value with a fixed number of existing failures, regardless of their spatial distribution in the communication graph.

\subsection{Most Vulnerable Sequence}
\begin{figure}
    \begin{subfigure}[t]{.45\linewidth}
        \centering
    	\includegraphics[width=\linewidth]{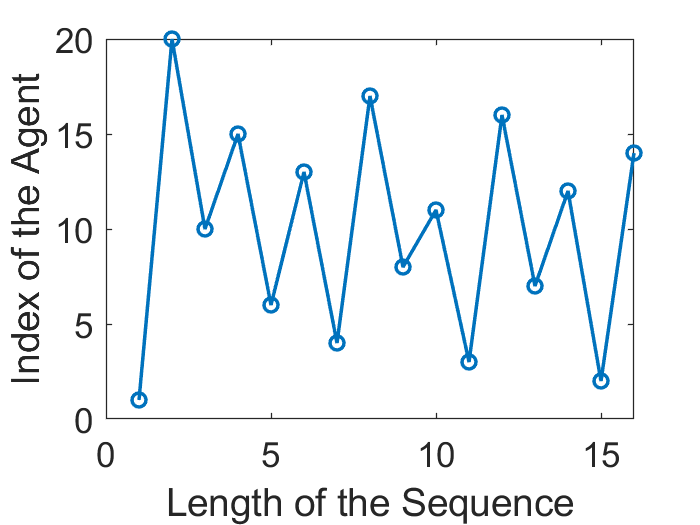}
    	\caption{The path graph.}
    \end{subfigure}
    \hfill
    \begin{subfigure}[t]{.45\linewidth}
        \centering
    	\includegraphics[width=\linewidth]{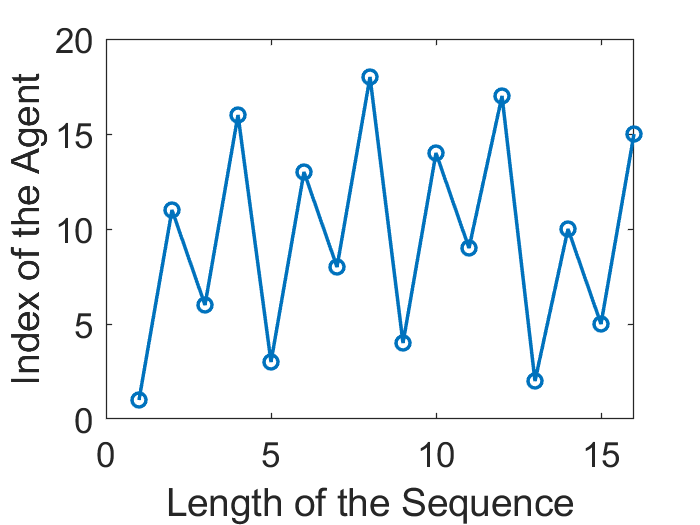}
    	\caption{The $1-$cycle graph.}
    \end{subfigure}
    \medskip
    \begin{subfigure}[t]{.45\linewidth}
        \centering
    	\includegraphics[width=\linewidth]{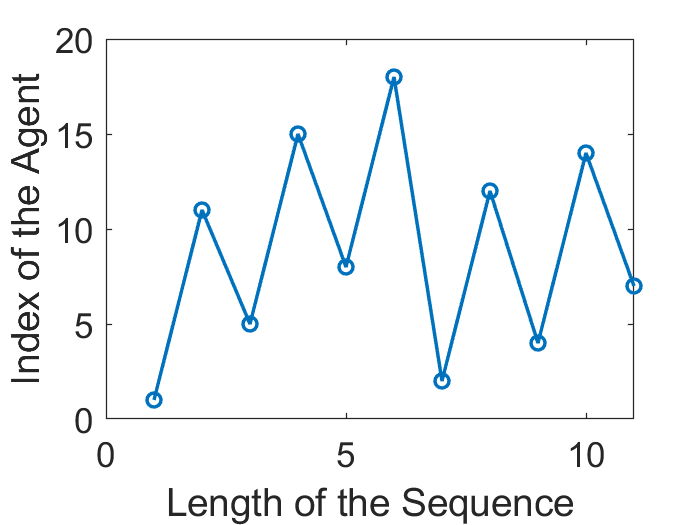}
    	\caption{The $5-$cycle graph.}
    \end{subfigure}
    \hfill
    \begin{subfigure}[t]{.45\linewidth}
        \centering
    	\includegraphics[width=\linewidth]{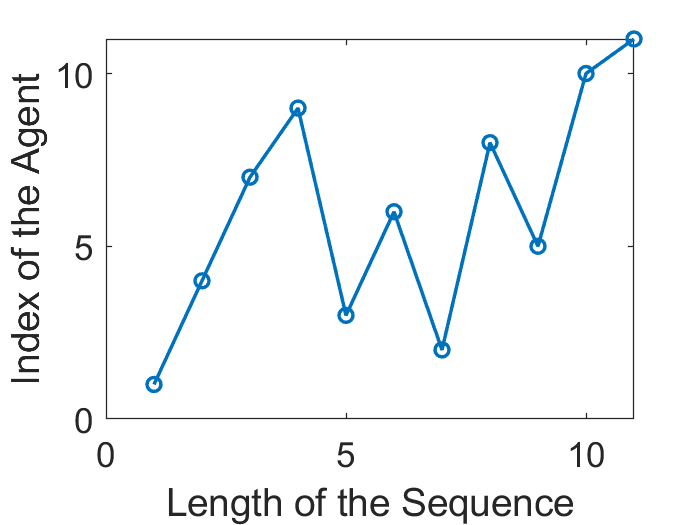}
    	\caption{The $8-$cycle graph.}
    \end{subfigure}
    \hfill
    \begin{subfigure}[t]{.45\linewidth}
        \centering
    	\includegraphics[width=\linewidth]{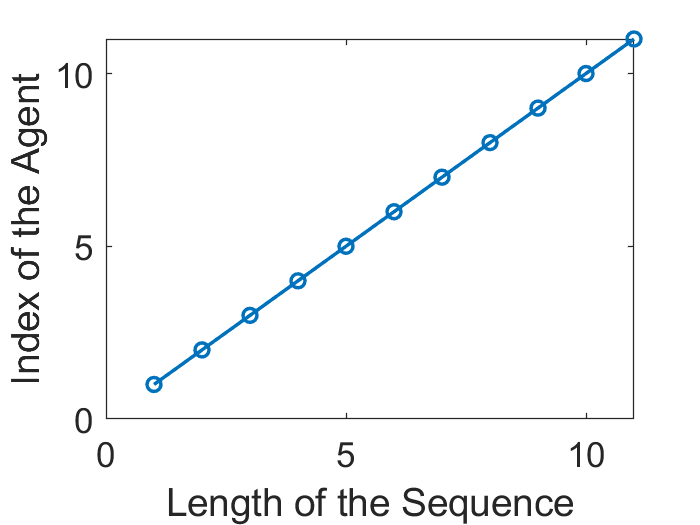}
    	\caption{The $9-$cycle graph.}
    \end{subfigure}
    \hfill
    \begin{subfigure}[t]{.45\linewidth}
        \centering
    	\includegraphics[width=\linewidth]{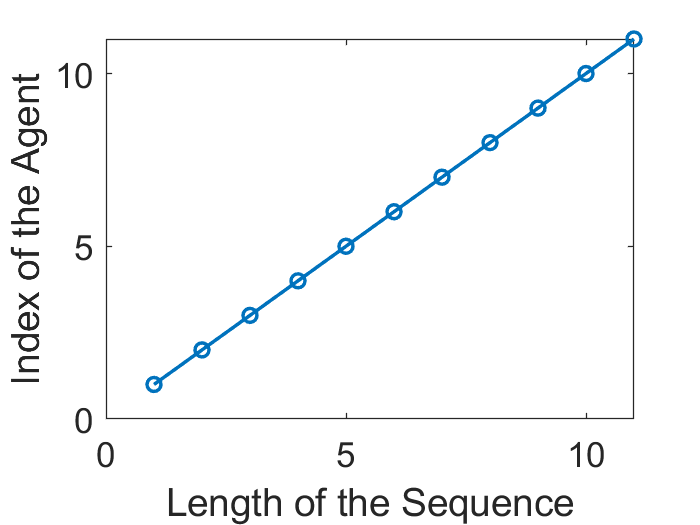}
    	\caption{The complete graph.}
    \end{subfigure}
    \caption{The most vulnerable sequence among various graphs.}
    \label{fig:risk_vulnerable}
\end{figure}
We also explore the emergence of the most vulnerable sequence of agents prone to cascading failures (Fig. \ref{fig:risk_vulnerable}). The most vulnerable sequence can be assessed by first finding the agent that obtains the highest risk and then finding the next most vulnerable agent by evaluating the cascading risk with the assumption that the previously found agents have failed to reach the $c-$consensus. The result in the complete graph indicates that any arbitrary sequence of agents is identically vulnerable since the agents will always obtain the same value of cascading risk in a complete graph. The investigation of the most vulnerable sequence provides information from the network design perspective such that one can adjust the systemic design based on the priority of the agents in the network.

%%%%%%%%%%%%%%%%%%%%%%%%%%%%%%%%%%%%%%%%%%%%%%%%%%%%%%%%%%%%%%%%%%%%%%%%%%%%%%%%%%%%%
\section{Conclusion}
To investigate the emergence of the cascading waves of fluctuations in the time-delay multi-agent systems, we develop a methodology to evaluate the risk of cascading failures using the notion of value-at-risk measure. The proposed cascading risk framework is a natural outgrowth of single event risk measure \cite{liu2022risk}, and it shows a significant advantage compared to the previous works as it can explore the scenarios when multiple system failures exist. In addition, to provide insights from the design perspective, we explore the time-delay induced hard limits on the best achievable levels of cascading risk as well as the update law for risk computation. Furthermore, the cascading risk framework reveals its advantages in the design of networks as it enables the designer to assess the sequence of the most vulnerable agents in a given network.

\appendix
\subsubsection{Proof of Lemma \ref{lem:y_steady}}
The result is an immediate extension of the steady-state statistics of the observables in \cite{Somarakis2019g} by considering a centering matrix $M_n$.
\hfill$\square$
    
\subsubsection{Proof of Lemma \ref{lem:multi-condition}}
The result follows directly after Lemma \ref{lem:y_steady} and the conditional distribution of a bi-variate normal random variable as in \cite{tong2012multivariate}. \hfill$\square$

\subsubsection{Proof of Theorem \ref{thm:cas_ren_risk}}
Considering Lemma \ref{lem:multi-condition}, one can rewrite the RHS of \eqref{eq:var} as 
$$
    \inf \big\{  \delta > 0  \big| \int_{-\infty}^{-\kappa_{\delta,+}^{\mathcal{I}_m,j}} e^{-t^2} dt +
    % &\hspace{2.5cm}
    \int_{-\infty}^{-\kappa_{\delta,-}^{\mathcal{I}_m,j}} e^{-t^2} dt < \sqrt{\pi} \varepsilon \big\}. 
$$
The result follows by taking cases on the risk value. More specifically, the case when $\casr = 0$ is equivalent to 
\begin{align*}
    \int_{-\infty}^{-\kappa_{0,+}^{\mathcal{I}_m,j}} e^{-t^2}\,dt + \int_{-\infty}^{-\kappa_{0,-}^{\mathcal{I}_m,j}} e^{-t^2}\,dt \leq \sqrt{\pi} \varepsilon,
\end{align*}
and the first branch condition follows from the definition of the error function. 
   
For the second branch, when $\casr \in (0,\infty]$, the monotonicity condition on $\kappa_{\delta,+}^{\mathcal{I}_m,j}$ and $\kappa_{\delta,-}^{\mathcal{I}_m,j}$ implies there is a unique $\delta^*>0$ such that 
$$
\int_{-\infty}^{-\kappa_{\delta^*,+}^{\mathcal{I}_m,j}} e^{-t^2}\,dt + \int_{-\infty}^{-\kappa_{\delta^*,-}^{\mathcal{I}_m,j}} e^{-t^2}\,dt=\sqrt{\pi}\varepsilon, 
$$ 
the result follows by solving the above equation for $\delta^*$. 
% \vivek{The integration limits in the above three integral should be checked. Probably, the left integral should look like $\int_{\kappa_{0,+}^{(i,j)}}^{\infty} e^{-t^2}  dt$} \chris{I agree. I think that you either use the $S(\delta)$ formulation (complimentary event , i.e. with reversed direction of inequality and $1-\varepsilon$) or change integration limits per Vivek's suggestion.}
\hfill$\square$

\subsubsection{Proof of Theorem \ref{thm:conditional_prob_update}}
To characterize the effect of the failures of $m + 1$ agents, let us consider the block covariance matrix
$$   
\tilde{\Sigma}_{22}' = \begin{bmatrix}
\tilde{\Sigma}_{22} & \tilde{\Sigma}_{21}(k)\\
\tilde{\Sigma}_{12}(k) & \tilde{\Sigma}_{kk}
\end{bmatrix},
$$
where $\tilde{\Sigma}_{kk} = \sigma _k^2, \tilde{\Sigma}_{21}(k) = \tilde{\Sigma}_{12}^T(k) =  \begin{bmatrix}
\sigma_{ki_1} & \dots  \sigma_{ki_m}
\end{bmatrix}$, and 
% $\tilde{\Sigma}_{22} = \begin{bmatrix}
% \sigma_{{k_1}{k_2}}
% \end{bmatrix}  \in \mathbb{R}^{m \times m}$ for all $k_1, k2 \in \mathcal{I}_m$.
$\tilde{\Sigma}_{22}$ is obtained from \eqref{eq:block_cov}. Since $\tilde{\Sigma}_{22}$ is invertible, we have 
\begin{align*}
    &\tilde{\Sigma}_{22}'^{-1} = \\
    &\hspace{1mm}\begin{bmatrix}
    \tilde{\Sigma}_{22}^{-1} (I_m+ \tilde{\Sigma}_{21}(k) S^{-1} \tilde{\Sigma}_{12}(k) \tilde{\Sigma}_{22}^{-1})  & - \tilde{\Sigma}_{22}^{-1}\tilde{\Sigma}_{21}(k)S^{-1}\\
    S^{-1} \tilde{\Sigma}_{12}(k) \tilde{\Sigma}_{22}^{-1} & S^{-1}
\end{bmatrix},
\end{align*}
where 
$
    S = \tilde{\Sigma}_{22}'/\tilde{\Sigma}_{22} =  \tilde{\Sigma}_{kk} -  \tilde{\Sigma}_{12}(k)\tilde{\Sigma}_{22}^{-1}\tilde{\Sigma}_{21}(k) = \tilde{\sigma}_k^2 
$
is the Schur complement \cite{diane1981schur} of block $\tilde{\Sigma}_{22}$ of the matrix $\tilde{\Sigma}_{22}'$. Let us consider the vector of failed observables of $(m+1)$ agents as
$
[
\bar{\bm{y}}_f ~ \bar{{y}}_{f_k}
]^T,$ where $\bm{\bar{y}}_{f} = [
\bar{y}_{f_1},...,\bar{y}_{f_m}
]^T$ is the vector of failed observables of $m$ agents and $\bar{{y}}_{f_k}$ is the failed observable of agent k, i.e., $(m+1)^{th}$ agent.
Consider the following vectors, 
$\tilde{\Sigma}_{12}' = [\tilde{\Sigma}_{12} ~ \tilde{\Sigma}_{12}(k)] = \tilde{\Sigma}_{12}'^{T}$ and the conditional cross-covariance of agents $j$ and $k$ after $m$ agents have failed $\tilde{\sigma}_{jk} = \sigma_{jk} - \tilde{\Sigma}_{12}(j) \tilde{\Sigma}_{22}^{-1} \tilde{\Sigma}_{21}(k)$,   
the result then follows directly by applying Lemma \ref{lem:multi-condition}. \hfill$\square$

%%%%%%%%%%%%%%%%%%%%%%%%%%%%%%%%%%%%%%%%%%%%%%%%%%%%%%%%%%%%%%%%%%%%%%%%%%%%%%%%%%%%%%%%%%%%%%
%END OF THE MAIN DOCUMENT
%%%%%%%%%%%%%%%%%%%%%%%%%%%%%%%%%%%%%%%%%%%%%%%%%%%%%%%%%%%%%%%%%%%%%%%%%%%%%%%%%%%%%%%%%%%%%%

\printbibliography

@book{van2010graph,
    title = {{Graph spectra for complex networks}},
    year = {2010},
    author = {Van Mieghem, Piet},
    publisher = {Cambridge University Press}
}

@book{Follmer2016,
    title = {{Stochastic Finance}},
    year = {2016},
    booktitle = {Stochastic Finance},
    author = {F{\"{o}}llmer, Hans and Schied, Alexander},
    month = {7},
    publisher = {De Gruyter}
}

@book{tong2012multivariate,
    title = {{The multivariate normal distribution}},
    year = {2012},
    author = {Tong, Yung Liang},
    publisher = {Springer Science {\&} Business Media}
}

@article{krueger1998perception,
  title={On the perception of social consensus},
  author={Krueger, J.},
  journal={Advances in experimental social psychology},
  volume={30},
  pages={163--240},
  year={1998},
  publisher={Elsevier}
}

@inproceedings{fagiolini2008consensus,
  title={Consensus-based distributed intrusion detection for multi-robot systems},
  author={Fagiolini, A. and Pellinacci, Marco and Valenti, Gianni and Dini, Gianluca and Bicchi, Antonio},
  booktitle={2008 IEEE International Conference on Robotics and Automation},
  pages={120--127},
  year={2008},
  organization={IEEE}
}

@ARTICLE{7438924,
  author={M. {Rahnamay-Naeini} and M. M. {Hayat}},
  journal={IEEE Transactions on Smart Grid}, 
  title={Cascading Failures in Interdependent Infrastructures: An Interdependent Markov-Chain Approach}, 
  year={2016},
  volume={7},
  number={4},
  pages={1997-2006},
  }

@article{zhang2019robustness,
  title={Robustness of interdependent cyber-physical systems against cascading failures},
  author={Zhang, Y. and Ya{\u{g}}an, O.},
  journal={IEEE Transactions on Automatic Control},
  volume={65},
  number={2},
  pages={711--726},
  year={2019},
  publisher={IEEE}
}

@article{zhang2018cascading,
  title={Cascading failures in interdependent systems under a flow redistribution model},
  author={Zhang, Y. and Arenas, A. and Ya{\u{g}}an, O.},
  journal={Physical Review E},
  volume={97},
  number={2},
  pages={022307},
  year={2018},
  publisher={APS}
}

@inproceedings{Somarakis2017a,
    title = {{Aggregate fluctuations in time-delay linear consensus networks: A systemic risk perspective}},
    year = {2017},
    booktitle = {Proceedings of the American Control Conference},
    author = {Somarakis, C. and Ghaedsharaf, Y. and Motee, N.}
}

@article{rockafellar2002conditional,
    title = {{Conditional value-at-risk for general loss distributions}},
    year = {2002},
    journal = {Journal of Banking and Finance},
    author = {Rockafellar, R. Tyrrell and Uryasev, Stanislav},
    number = {7},
    pages = {1443--1471},
    volume = {26}
}

@inproceedings{Somarakis2016g,
    title = {{Interplays Between Systemic Risk and Network Topology in Consensus Networks}},
    year = {2016},
    booktitle = {IFAC-PapersOnLine},
    author = {Somarakis, C. and Siami, M. and Motee, N.},
    number = {22},
    volume = {49}
}

@article{rockafellar2000optimization,
    title = {{Optimization of Conditional Value-at-Risk}},
    year = {1999},
    journal = {Portfolio The Magazine Of The Fine Arts},
    author = {Rockafellar, R. T. and Uryasev, S.},
    pages = {1--26},
    volume = {2}
}

@article{Somarakis2020b,
    title = {{Risk of Collision and Detachment in Vehicle Platooning: Time-Delay-Induced Limitations and Tradeoffs}},
    year = {2020},
    journal = {IEEE Transactions on Automatic Control},
    author = {Somarakis, C. and Ghaedsharaf, Y. and Motee, N.},
    number = {8},
    volume = {65}
}

@article{Somarakis2019g,
    title = {{Time-delay origins of fundamental tradeoffs between risk of large fluctuations and network connectivity}},
    year = {2019},
    journal = {IEEE Transactions on Automatic Control},
    author = {Somarakis, C. and Ghaedsharaf, Y. and Motee, N.},
    number = {9},
    volume = {64}
}

@inproceedings{liu2021risk,
    title={Risk of Cascading Failures in Time-Delayed Vehicle Platooning}, 
    author={G. Liu and C. Somarakis and N. Motee},
    year={2021},
    booktitle = {IEEE Conference on Decision and Control},
    % eprint={2109.01963},
    % archivePrefix={arXiv},
}

@article{artzner1997thinking,
  title={Thinking coherently},
  author={Artzner, P.},
  journal={Risk},
  pages={68--71},
  year={1997}
}

@article{artzner1999coherent,
  title={Coherent measures of risk},
  author={Artzner, P. and Delbaen, F. and Eber, J. and Heath, D.},
  journal={Mathematical finance},
  volume={9},
  number={3},
  pages={203--228},
  year={1999},
  publisher={Wiley Online Library}
}

@article{ren2007information,
  title={Information consensus in multivehicle cooperative control},
  author={Ren, W. and Beard, R. W. and Atkins, E. M.},
  journal={IEEE Control systems magazine},
  volume={27},
  number={2},
  pages={71--82},
  year={2007},
  publisher={IEEE}
}

@article{olfati2007consensus,
  title={Consensus and cooperation in networked multi-agent systems},
  author={Olfati-Saber, R. and Fax, J. A. and Murray, R. M.},
  journal={Proceedings of the IEEE},
  volume={95},
  number={1},
  pages={215--233},
  year={2007},
  publisher={IEEE}
}

@article{olfati2004consensus,
  title={Consensus problems in networks of agents with switching topology and time-delays},
  author={Olfati-Saber, R. and Murray, R. M.},
  journal={IEEE Transactions on automatic control},
  volume={49},
  number={9},
  pages={1520--1533},
  year={2004},
  publisher={IEEE}
}

@article{diane1981schur,
  title={Schur Complements and Statistics},
  author={Diane Valerie Ouellette},
  journal={Linear Algebra and Its Applications},
  volume={36},
  number={9},
  pages={187--295},
  year={1981},
  publisher={Elsevier}
}

@inproceedings{liu2022risk,
  title={Risk of Cascading Failures in Multi-agent Rendezvous with Communication Time Delay},
  author={Liu, Guangyi and Pandey, Vivek and Somarakis, Christoforos and Motee, Nader},
  booktitle={2022 American Control Conference (ACC)},
  pages={2172--2177},
  year={2022},
  organization={IEEE}
}

@inproceedings{bragagnolo2019attack,
  title={Attack detection in a cluster divided consensus network},
  author={Bragagnolo, Marcos Cesar and Messai, Nadhir and Manamanni, Noureddine},
  booktitle={2019 18th European Control Conference (ECC)},
  pages={1091--1096},
  year={2019},
  organization={IEEE}
}

@inproceedings{slay2007lessons,
  title={Lessons learned from the maroochy water breach},
  author={Slay, Jill and Miller, Michael},
  booktitle={International conference on critical infrastructure protection},
  pages={73--82},
  year={2007},
  organization={Springer}
}

@article{kushner2013real,
  title={The real story of stuxnet},
  author={Kushner, David},
  journal={ieee Spectrum},
  volume={50},
  number={3},
  pages={48--53},
  year={2013},
  publisher={Institute of Electrical and Electronics Engineers, Inc., 345 E. 47 th St. NY~…}
}

@inproceedings{liu2022emergence,
  title={Emergence of Cascading Risk and Role of Spatial Locations of Collisions in Time-Delayed Platoon of Vehicles},
  author={Liu, Guangyi and Somarakis, Christoforos and Motee, Nader},
  booktitle={2022 IEEE 61st Conference on Decision and Control (CDC)},
  pages={6460--6465},
  year={2022},
  organization={IEEE}
}
\end{document}